\documentclass[structabstract]{aa}  
\usepackage{epsfig ,graphicx}
\usepackage{txfonts}
\usepackage{natbib}
\bibpunct{(}{)}{;}{a}{}{,} 
\begin{document}

   \title{Constraining the Circumbinary Envelope of Z CMa via imaging polarimetry
    \thanks{Based on observations made with the \textit{William Herschel} Telescope operated on the island of La Palma by the Isaac Newton Group in the Spanish Observatorio del Roque de los Muchachos of the Instituto de Astrof\'isica de Canarias}}


   \author{H. Canovas\inst{1,5},
   M. Min\inst{2,5}, S. V. Jeffers\inst{3,5}, M. Rodenhuis\inst{4,5}, C. U. Keller\inst{4,5}
        }

   \institute{Departamento de Fisica y Astronomia, Universidad de Valpara\' iso, Valpara\' iso, Chile\\
   \email{H.CanovasCabrera@uu.nl}
   \and
   Astronomical Institute Anton Pannekoek, University of Amsterdam,Kruislaan 403, 1098 SJ Amsterdam, The Netherlands
   \and
   Institut fur Astrophysik Goettingen, Friedrich-Hund-Platz 1, 37077 Goettingen
   \and
   Leiden Observatory, Leiden University, P.O. Box 9513, NL-2300 RA Leiden, The Netherlands
   \and
   Sterrekundig Instituut, Universiteit Utrecht, PO Box 80000, 3508 TA Utrecht, The Netherlands
                }

\date{Received December xx, 2011}

    \abstract
     {Z CMa is a complex binary system, composed of a Herbig Be and an FU Ori star. The Herbig star is surrounded by a dust cocoon of variable geometry, and the whole
   system is surrounded by an infalling envelope. Previous spectropolarimetric observations have reported a preferred orientation of the polarization angle, perpendicular
   to the direction of a large, parsec-sized jet associated with the Herbig star.}
   {The variability in the amount of polarized light has been associated to changes in the geometry of the dust cocoon that surrounds the Herbig star.
   We aim to constrain the properties of Z CMa by means of imaging polarimetry at optical wavelengths.}
   {Using ExPo, a dual-beam imaging polarimeter which operates at optical wavelengths, we have obtained imaging (linear) polarimetric data of Z CMa. Our observations were 
   secured during the return to quiescence after the 2008 outburst.}
     {We detect three polarized features over Z CMa. Two of these features are related to the two jets reported in this system: the large jet associated to the Herbig star,
     and the micro-jet associated to the FU Ori star. Our results suggest that the micro-jet extends to a distance ten times larger than reported in previous studies.
     The third feature suggests the presence of a hole in the dust cocoon that surrounds the Herbig star of this system.
   According to our simulations, this hole can produce a pencil beam of light that we see scattered off the low-density envelope surrounding the system.}
 {}
   \keywords{circumstellar matter -- stars: variables: T Tauri -- herbig Ae/Be -- stars: winds -- outflows -- stars: individual: Z CMa -- techniques: polarimetry}

\titlerunning{Constraining the Circumbinary Envelope of Z CMa via imaging polarimetry}
\authorrunning{Canovas et al.}
\maketitle

\section{Introduction}
Z CMa is one of the most complex young binary systems known to date. Originally identified as an FU Ori star \citep{Hartmann_1989}, 
recent studies show a much more complicated scenario where a Herbig Be star inside a dust cocoon cohabits with  an FU Ori star and an in-falling envelope 
surrounding both stars \citep{Alonso_2009}. Furthermore, a 3.6 parsec-sized jet and  a micro-jet are associated to the Herbig and the FU Ori 
stars, respectively \citep{Whelan_2010}. Z CMa belongs to the CMa OB1 association, with distances estimates ranging 
from 930 pc to 1150 pc \citep{Claria_1974,Herbst_1978,Ibragimov_1990,Kaltcheva_2000}.

Speckle interferometry at near-infrared \citep{Koresko_1991,Haas_1993} and optical \citep{Barth_1994,Thiebaut_1995} wavelengths revealed a companion to
the FU Ori star, placed at 0.1" from it, and with a position angle of $305^{\circ} \pm 2^{\circ}$, East of North. This object showed an excess of flux at
infrared \citep[J, H and K filters, ][]{Koresko_1991} and millimeter \citep[$\lambda = 1.1$ mm, ][]{Beckwith_1991} wavelengths, which suggested the presence
of a dust distribution around it. A stellar photosphere or an accretion disk could not explain this cool excess because of the youth of the system \citep{Koresko_1991}. 
Spectropolarimetric measurements of Z CMa \citep{Whitney_1993} confirmed the existence of an asymmetrical, 
geometrically variable, dust cocoon around the infrared, more massive, Herbig-like star (hereafter \textit{primary}).

The presence of circumbinary and circumstellar disks has been investigated by several authors. \citet{Malbet_1993} reported the detection of a circumbinary
disk centered on the FU Ori star, but independent observations could not confirm the existence of such a disk  \citep{Tessier_1994}. Speckle polarimetry
at mid-infrared wavelengths showed that both stars of the Z CMa binary are polarized at these wavelengths, which could be explained by means of an inclined
disk around the secondary \citep{Fischer_1998}. Observations at millimeter wavelengths indicate the presence of an inclined toroid of cold dust with an inner
radius of 2000 AU and and outer radius of 5000 AU, but a spherical envelope can explain better the extinction of this system \citep{Alonso_2009}.

The large jet associated to Z CMa has a total length of 3.6 pc \citep{Poetzel_1989}.
This jet has been associated to the primary \citep{Garcia_1999} and to the secondary \citep{Velazques_2001}. A recent  analysis of the jet properties
shows that there are actually two jets produced by this system: the pc sized one, associated to the primary, with a position angle of $245^{\circ}$,
and a micro-jet, associated to the secondary, with a position angle of $235^{\circ}$ \citep{Whelan_2010}. The latter has been measured up to
a distance of 0.4" from the FU Ori star.

\begin{figure}[h]
  \center
   \includegraphics[width = 1\linewidth,trim = 20 20 0 0]{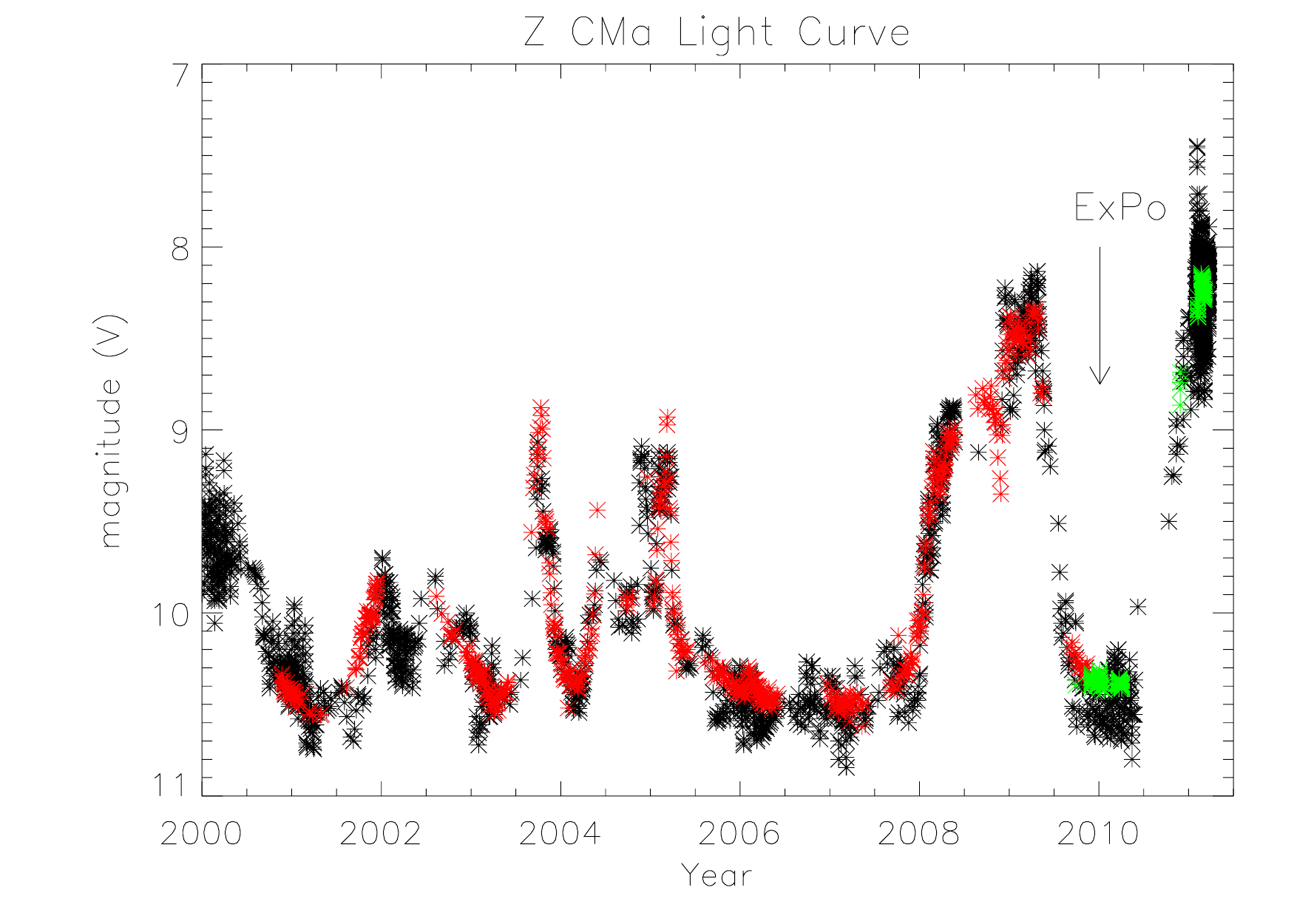}
   \caption{Light Curve of Z CMa. Black and red asterisks correspond to data from the American Association of Variable Star Observers (AAVSO) and the All Sky Automated Survey 
   (ASAS) \citep{Pojmanski_2002}, respectively. Green asterisks correspond to data from the Czech Astronomical Society (CAS).}
  \label{fig:zcma_fig2}
\end{figure}

In 2008 Z CMa suffered the strongest outburst ever reported for this object. The nature of this outburst is still under debate, and it is not clear wether it
was originated as a consequence of a real outburst associated to the primary \citep{Benisty_2010}, or because of changes (such as the formation of a
new hole) in the dust cocoon \citep{Szeifert_2010}.
\begin{figure}[h]
  \center
   \includegraphics[width = 1\linewidth,trim = 0 0 0 0]{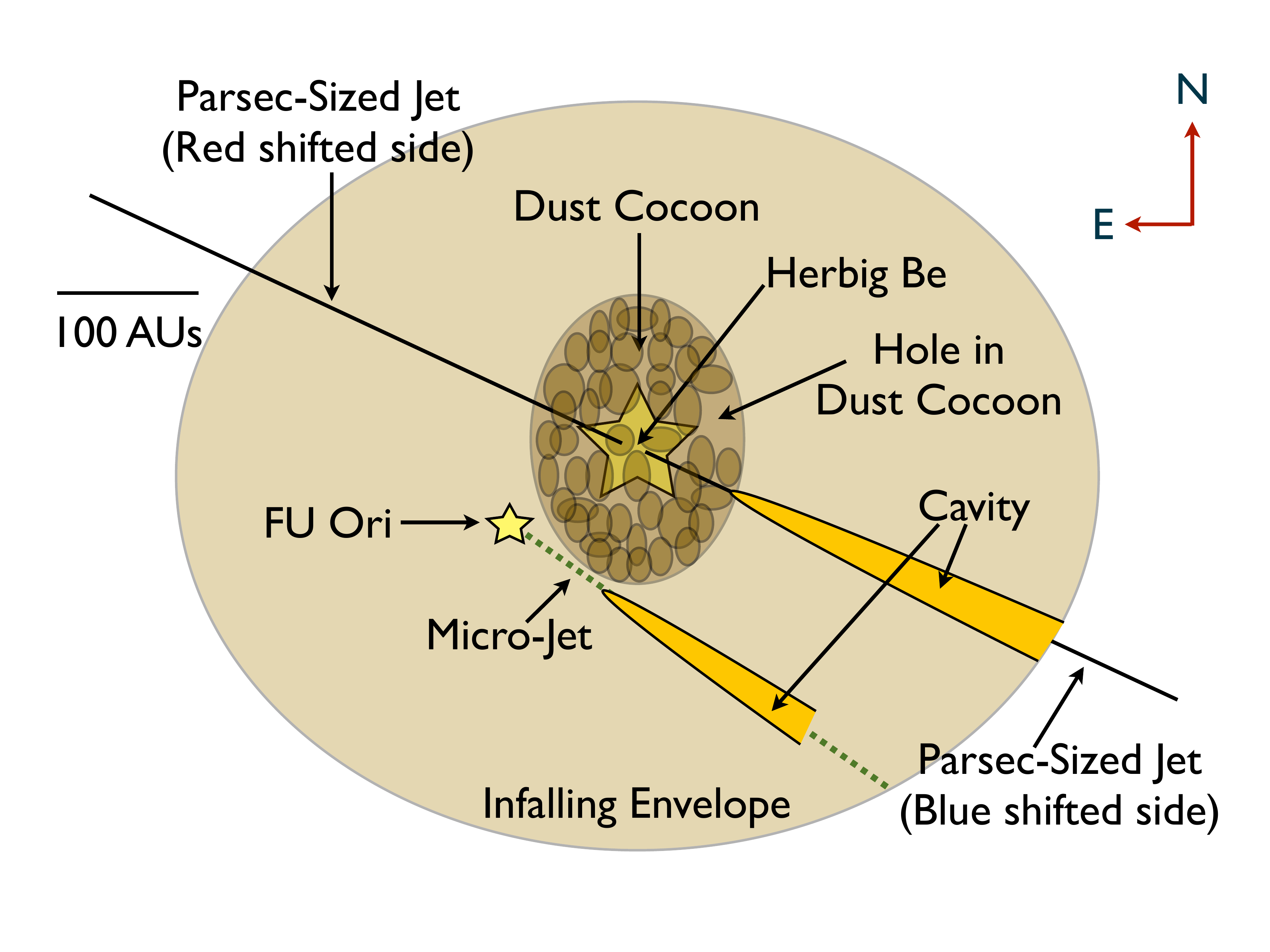}
   \caption{Schematic (non scaled) picture of the Z CMa system, as seen from Earth. The two stars are separated by $\sim0.1"$. (100 AUs assuming a distance to 
   Z CMa of 1150 pc).  The large jet associated to the primary, with an orientation of $245^{\circ}$, is indicated by the black line, while the micro-jet, with an inclination
   of $235^{\circ}$, is indicated by the green dashed line.}
  \label{fig:zcma_fig1}
\end{figure}
A very simplified scheme of the current picture of Z CMa is shown in Fig.~\ref{fig:zcma_fig1}. Currently Z CMa is experiencing a new, stronger outburst,
which started mid 2010 and lasts until current date (see Fig.~\ref{fig:zcma_fig2}).

The complex nature of this system makes imaging polarimetry a very interesting tool to better characterize it. While several authors have 
performed imaging polarimetry studies of circumstellar environments around young stars \citep[e.g. ][]{Breger_1973,Bastien_1988,Close_1997,
Kuhn_2001,Oppenheimer_2008,Perrin_2009_MWC778}, there are no imaging polarimetric observations of Z CMa to date.
In this paper we present imaging polarimetry at optical wavelengths of Z CMa  obtained with ExPo, the Extreme Polarimeter \citep[see][]{Keller_2006,Rodenhuis_2008},
currently a visitor instrument at the 4.2 meters William Herschel Telescope (WHT). We describe the observations, the instrument and the
data reduction in Sect. 2. The image analysis and models are discussed in Sect. 3. The discussion and conclusion are written in Sect. 4 and Sect. 5, respectively.

\section{Observations and Data Processing}

Z CMa was observed on the 30th of December, 2009 (Night 1) and the 4th of January, 2010 (Night 2), during the third campaign ($27/12/2009-04/01/2010$) of ExPo as a visitor instrument 
at the WHT. The seeing during these two nights was fairly bad, with an average value of  $\sim1.47"$. An unpolarized, diskless star (G 91-23) was observed the night of the 31st of December, 
2009 for comparison purposes. No filter was used during all these measurements, so the full optical range was covered (ExPo is sensitive to the wavelength range from 400 to 
900 nanometers). Table~\ref{tab:observations} summarizes these observations. A set of calibration flat-fields was taken at the beginning and the end of each night, and a set of dark frames
 was taken at the beginning of each observation.
\begin{table}
		\centering
				\caption{Summary of ExPo observations.}
			\begin{tabular}{ c  c  c  c c }
\hline\hline
Target			&	Date				&	No Exp.				& Total Exp. Time	&	Seeing	\\
				&						&					& [s]				&	["]		\\
\hline
Z CMa			&	30 Dec 2009		&	$4\times18000$		&2016			&	1.51		\\
				&	04 Jan 2010		&	$4\times20600$		&2307			&	1.44		\\
G 91-23			&	31 Dec 2009		&	$4\times4095$			&458.64			&	0.95			\\
\hline
			\end{tabular}
		\tablefoot{ The exposure time for each single frame was 0.028 seconds in all cases. No filter was used in any of these observations. G 91-23 is an
		unpolarized, diskless star.}
	\label{tab:observations}
\end{table}
\subsection{Instrumental Description}

ExPo is a dual-beam imaging polarimeter working at optical wavelengths.
It combines a fast modulating ferroelectric liquid crystal (FLC), a cube beamsplitter (BS) and an Electron Multiplying Charge Coupled Device (EM-CCD).
The FLC modulates the polarization state of the incoming light by $90^{\circ}$ every 0.028 seconds, switching between two states ``A" and ``B". 
By changing the voltage applied, the FLC modifies its state, without having to rotate any part of the instrument.\footnote{The FLC can actually
modify its state at much higher rates than the 35 Hz used by ExPo. The exposure time in this instrument is currently limited by the EM-CCD, which is
unable to operate at shorter exposure times than 0.028 seconds.} The beamsplitter divides 
the incoming light into two beams (\textit{left} and \textit{right} beams) with orthogonal polarization states, which are then imaged onto two different regions
of the $512\times512$ pixels EM-CCD. The current field of view of this instrument is $20"\times20"$, which is projected onto an area of the CCD of  $256\times256$ pixels,
resulting in a pixel size of 0.078"/px.

At the end of one FLC cycle, four different images, $A_\mathrm{left},A_\mathrm{right},B_\mathrm{left}, B_\mathrm{right}$ are produced. By subtracting the two
 simultaneous images recorded at the ``A" and  ``B" states a ``difference", linearly polarized image, is produced:
\begin{equation}
\Delta A = A_\mathrm{left} - A_\mathrm{right},
\label{eq:deltaA}
\end{equation}
\begin{equation}
\Delta B =  B_\mathrm{left} - B_\mathrm{right}.
\label{eq:deltaB}
\end{equation}
A set of narrowband and broadband filters is implemented by a filter wheel, though none of these filters were used in the observations presented here.
A standard ExPo observation comprises at least 4095 images ($\sim$2 minutes) recorded at
one fixed FLC position. Once a set of measurements is finished, the FLC is rotated by $22.5^{\circ}$, and the procedure is repeated. At the end of the observation,
four sets of measurements with the FLC oriented at $0^{\circ},22.5^{\circ},45^{\circ}$, and $67.5^{\circ}$ are recorded during each observation.
\subsection{Data Analysis}

Each one of the four images produced after one FLC cycle is finished are individually corrected of dark, bias, flat-field and cosmic rays.
These images are then aligned by means of a cross-correlation algorithm. To increase the accuracy of this process, two different
point spread functions (PSFs) templates, one for the left beam, and the other for the right beam, are produced during the reduction. 
The $A_\mathrm{left}$ and $B_\mathrm{left}$ images are then aligned accordingly to the coordinates of the maximum of the cross-correlation
of each image with the left-beam template. The same process is repeated with the right-beam images and the right-beam template.
For  a detailed explanation of the data reduction techniques applied to the ExPo instrument, see \citet{Canovas_2011}.
The short exposure time used by ExPo (0.028 seconds) allows us to minimize the tip-tilt error of the wavefront by properly aligning our images.
The final full width half maximum (FWHM) of our average PSF after centering  is then $\sim 1.2"$, which supposes an improvement of about
$\sim 20\%$ with respect to the average seeing during our observations. A polarization image is obtained from a ``double-difference" \citep{Kuhn_2001},
by combining Eq.~\ref{eq:deltaA} and Eq.~\ref{eq:deltaB}:
\begin{equation}
P_\mathrm{I}^{'} = 0.5 \left(\Delta A - \Delta B \right) = 0.5 \left((A_\mathrm{left} - A_\mathrm{right}) - (B_\mathrm{left} - B_\mathrm{right}) \right).
\label{eq:eq3}
\end{equation}
$P_\mathrm{I}^{'}$ is a polarized, yet uncalibrated, image. 
Because of the fast modulation of the FLC and the combination of two simultaneous images with opposite polarization states, most of the instrument systematic
errors such as the highly static speckle noise \citep[see, for instance,][]{Racine_1999,Sivaramakrishnan_2002,Soummer_2007} are removed \citep[see for example][]
{Hinkley_2009}, allowing us to increase our sensitivity to faint, polarized features. The intensity image is calculated as the sum of the four images: 
\begin{equation}
I =  0.5 \left(A_\mathrm{left} + A_\mathrm{right} + B_\mathrm{left} + B_\mathrm{right} \right).
\label{eq:eq4}
\end{equation}
The instrumental polarization is removed by subtracting the average polarization degree in a circular area of radius 5 pixels ($0.39"$), centered in the star,
to the polarized images \citep[for a detailed explanation of this process and its results when applied to ExPo see][]{Canovas_2011,Min_2012}.
By doing this, we remove most of the instrumental polarization caused by the elements of the telescope + instrument system, but we also
remove polarized light from the central star, in case the star itself is polarized. This method has been applied before by several authors
\citep[see for example][]{Perrin_2008,Quanz_2011} when observing with different imaging polarimeters. The main advantage of this technique
is that it allows us to trace the spatial variations in polarized intensity with high accuracy. As a collateral effect, the
polarization degree measured with this instrument provides us with a lower limit of the true polarization degree in the observed targets.

Because ExPo does not have a de-rotator, sky rotation must be corrected during the data reduction. To properly correct this, not only a physical rotation of the image is needed
(to preserve the north orientation when combining different datasets), but also a vectorial rotation of the Stokes Q and U images must be performed. After correcting this, the Stokes
parameters Q and U can be calculated in a reference frame fixed to the observed target. The sky polarization is corrected once the calibrated Stokes Q and U images are produced.
To do this, the polarization of the sky is computed as the median of four different sky regions in the calibrated Stokes Q images. This value is then subtracted from these images,
and the same process is repeated with the Stokes U images.

We produced two calibrated datasets from each observation. To produce the Set 1 we combined and calibrated the images obtained with the
FLC rotated at  $0^{\circ}$ and $22.5^{\circ}$, while Set 2 was produced by using the images obtained with the FLC rotated at $45^{\circ}$ and
$67.5^{\circ}$. Since in each observation the total amount of images observed in each one of the four FLC position is the same (see Table~\ref{tab:observations})
both Set 1 and Set 2 have the same total exposure time. The polarized intensity (linearly polarized light) is calculated as $P_\mathrm{I} = \sqrt{Q^{2} + U^{2}}$, and the 
degree of polarization as $P = P_\mathrm{I}/I$, where $I$ is the total intensity. The polarization angle, $P_\mathrm{\Theta} = 0.5 \cdot \mathrm{arctan}\left(\frac{U}{Q}\right)$ 
defines the orientation of the polarization plane. The error on the polarization angle depends on the error of the calibration process, but also on the region of the image where it 
is calculated. To minimize errors, the stokes Q and U images are binned with a $5\times5$ pixel (i.e. $0.39"\times0.39"$) box before $P_\mathrm{\Theta}$ is computed. 
Furthermore, $P_\mathrm{\Theta}$ is calculated 
only in the regions of the image with a signal-to-noise ratio (SNR) higher than $3$ to reduce the error introduced by the noise. The accuracy of $P_\mathrm{\Theta}$ depends then
on the position of the image where it is calculated. For Z CMa, the maximum error that we obtain for $P_\mathrm{\Theta}$ is of the order of $\pm 3.8^{\circ}$ in the regions with SNR above 3.
\begin{figure*}
  \center
   \includegraphics[width = 1.\linewidth,trim = 0 40 30 90]{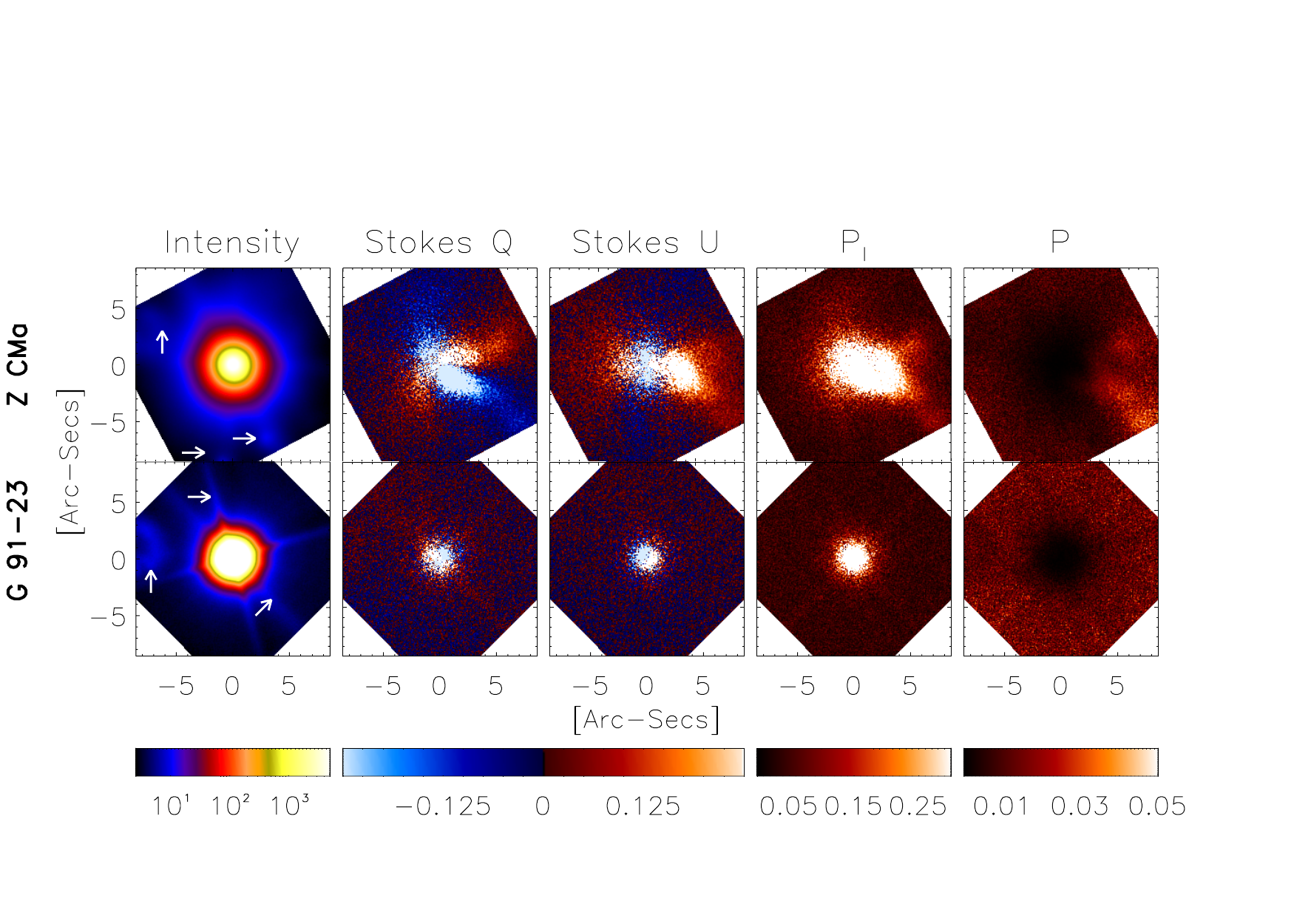}
   \caption{Comparison of Z CMa (top row) and G 91-23 (unpolarized, diskless star, bottom row). North is up and East is left in all the images. Starting from the left, the
   first column shows the total intensity image in logarithmic scale. The arrows indicate the position of different instrumental artifacts. The second and third column
   show the Stokes Q and U parameters, respectively. The polarized intensity ($P_\mathrm{I}$) is shown in the fourth column. G 91-23 shows a pattern caused by
   remnant, uncorrected, noise. Z CMa shows an extended, asymmetrical structure, with two extended features towards West, and South-West. The fifth column shows
   the degree of polarization ($P$). The units of the color bars are given in CCD-counts, except in the case of $P$, where the degree ($\%$) of polarization is shown.
   Our polarized images are calibrated in a reference system fixed to the observed object (see Sect. 2).}
  \label{fig:zcma_fig3}
\end{figure*}
\begin{figure*}[!h]
  \center
      \includegraphics[width = 0.8\linewidth,trim = 60 140 40 20]{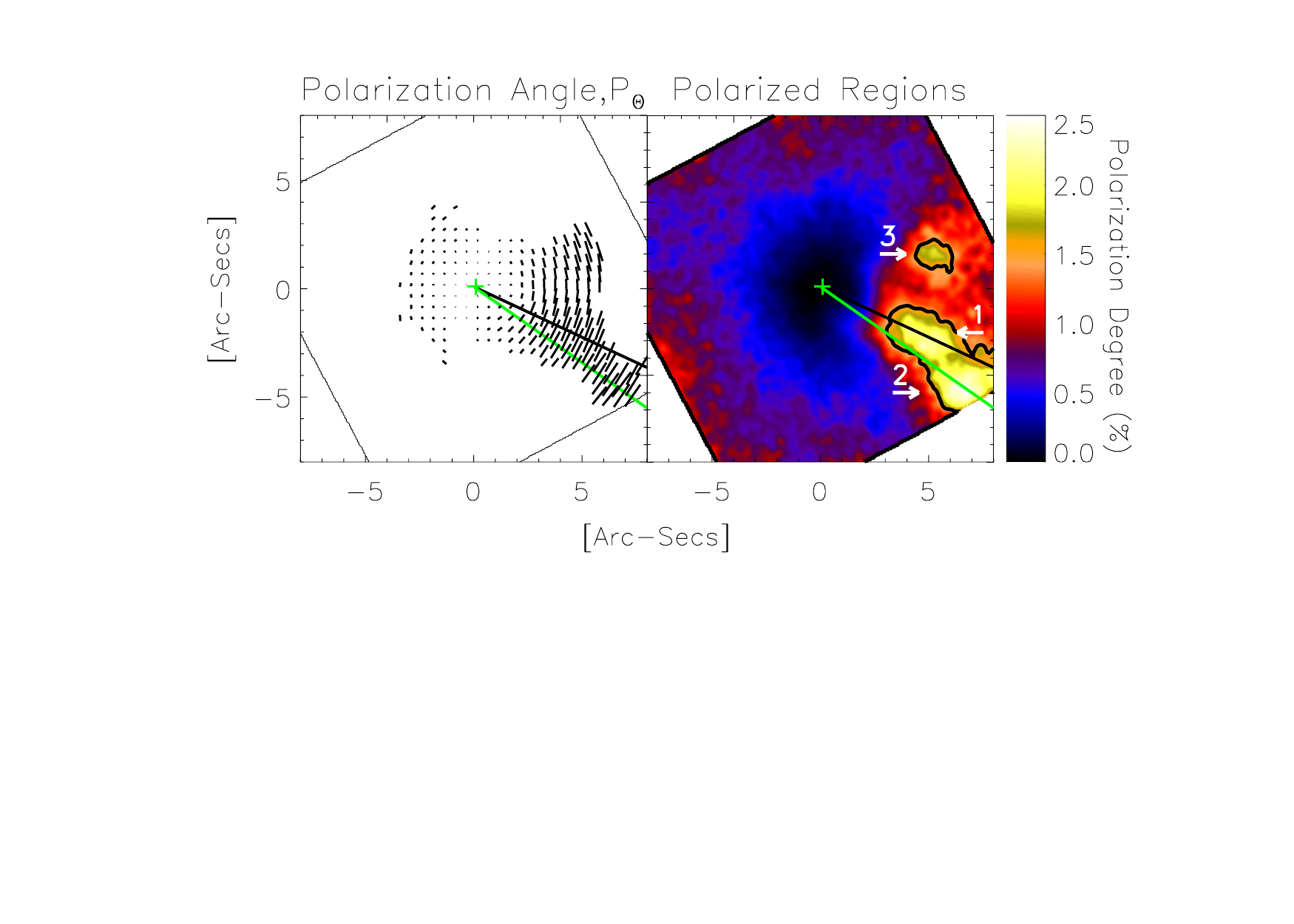}
   \caption{Degree of polarization of Z CMa.  North is up, East id left. The black and green lines show the trajectories of the large jet and micro-jet, respectively.
   The green cross shows the position of the binary; the separation of the stars is smaller than the size of the symbol. Left: $P_\mathrm{\Theta}$ is shown by the
   vectors. Only regions of the image with SNR higher than 3 are displayed in this image. The length of the vectors is proportional to the local degree of polarization.
   Right: $P$ image smoothed by convolution with a gaussian kernel of 5 pixels. The black contours indicate the regions of the image where $P$ is higher than
   $1.5\%$. The three arrows indicate the three regions of the image discussed in Sect. 4.}
    \label{fig:zcma_fig5}
\end{figure*}
\section{Results}
A total amount of four calibrated datasets for Z CMa (two per night), and two calibrated datasets for the diskless, unpolarized star were obtained. 
Data from Night 2 show a better SNR than data from Night 1: the Moon was closer to Z CMa in Night 1, increasing the amount of sky background
noise. Therefore, we will focus our analysis on the first of the two datasets obtained during Night 2, which is the set with higher SNR. 

Figure~\ref{fig:zcma_fig3} shows a comparison between Z CMa and a diskless, unpolarized star (G 91-23). The first column, starting from the left side,shows
the intensity image in logarithmic scale. The color scale in this case is chosen to enhance the instrumental artifacts in our images. In both images, the vertical arrow
indicates the position of the ghost caused by the beamsplitter. The two horizontal arrows in the Z CMa image show the position of two ghosts caused by the FLC.
The horizontal arrow in G 91-23 shows the position of one of the spiders of the telescope. The pattern produced by these spiders is clearly visible in this image.
The inclined arrow indicates the effect of a non-perfect correction of the CCD-smearing effect. The two components of Z CMa appear unresolved in our images.
The second and third columns show the stokes Q and U parameters, respectively. In the case of Z CMa, an extended, butterfly-like pattern is evident, while in the case of 
G 91-23 these two images are dominated by remnant noise. The fourth column shows the polarized intensity, $P_\mathrm{I}$, in which a very different pattern for each star is also clear. 
In the case of G 91-23, the $P_\mathrm{I}$ image is dominated by the remnant noise that appears in the Q and U images. The noise from these images adds quadratically 
when computing $P_\mathrm{I}$, producing the observed pattern. In the case of Z CMa, $P_\mathrm{I}$ shows a very strong asymmetry, with two extended features towards the
West direction (to the right side of the image), and the South-West (to the bottom-right corner of the image). Because of the small separation of the two stars, it is
impossible to resolve the dust cocoon around the primary in our images. Therefore, the extended polarized pattern that we see in Z CMa is probably caused by
the circumbinary disk and infalling envelope. The fifth column shows the degree of polarization ($P = P_\mathrm{I}/I$) of Z CMa and the comparison star. As it happens
with the polarized intensity, the unpolarized star does not show any  spatial structure but a centrosymmetric pattern. The polarized intensity, $P_\mathrm{I}$, which is by
definition a positive quantity, is biased upwards by the noise. On the other hand, the intensity $I$ approaches zero when getting close the edge of the image.
These two effects combined produce what we see in the $P$ image of the unpolarized star, where $P$ increases its value when approaching the image's edges.
The image of Z CMa, however, shows again two polarization features, oriented to West and South-West, respectively. 

Further analysis shows that these two polarized structures appear at the same position in all the observations of Z CMa. Z CMa 
was observed at a different altitude during Night 2 when compared to Night 1, but in all images we detect the same polarization features. In case these two polarized
features were caused by internal reflections or another instrumental artifact, they should appear in different positions when comparing images from Night 1 with images
from Night 2. This rules out the possibility that these two features are caused by any instrumental or data processing artifact. \

To reduce the noise in the polarization degree image, we have convolved the best dataset of our observations (Night 2, Set 1) with a gaussian kernel with FWHM
of 5 pixels ($0.39"$). The resulting image is shown in Fig~\ref{fig:zcma_fig5}.
The left side of Fig~\ref{fig:zcma_fig5} shows $P_\mathrm{\Theta}$. Only regions of the image with a SNR higher than 3 are displayed on this figure. The trajectory of the large
jet (blue-shifted component) associated to the Herbig star is indicated by the black line, at a position angle of $245^{\circ}$.
The green line shows the position of the micro-jet associated to the FU Ori star, at a position angle of $235^{\circ}$ \citep[see][]{Whelan_2010}.
The position of the primary and the secondary is marked by the green cross at the center of the image.
The right side of Fig.~\ref{fig:zcma_fig5} shows the polarization degree image. The color scale is chosen to
enhance the polarized features that appear in this image. The West side of Z CMa shows a polarized continuum with a degree of polarization of about 1.2\% (the red
colored regions). The orange and yellow regions indicated by the three arrows have a polarization degree higher than 1.5\% The mean and  standard deviation,
$\sigma$, of $P$ and $P_\mathrm{\Theta}$ of these three regions is shown in Table~\ref{tab:tab2}, together with results from previous spectropolarimetric observations of Z CMa.
\begin{table}[h]
	\centering
		\caption{Previous spectropolarimetric measurements of Z CMa and ExPo results for the three regions shown in Fig.~\ref{fig:zcma_fig5}} 
		\begin{tabular}{c c c c}
\hline\hline
Date						&	$\lambda$ Range				&	$P$					&	$P_\mathrm{\Theta}$				\\
$[\mathrm{year}]$			&	 [nm]							&	$[\%]$				&	$[^{\circ}]$							\\
\hline
1991\tablefootmark{1}		&	412-687						&	$2.00\pm0.005$		&	$150\pm1.5$							\\	
1992\tablefootmark{1}		&	450-620						&	$1.40\pm0.005$		&	$154\pm1.5$							\\	
2008\tablefootmark{2}		&	331-920						&	$2.6\pm0.1$			&	$160\pm1$							\\
2010\tablefootmark{3}		&	400-900						&	$1.9\pm0.1$			&	$155.4\pm3.9$						\\
2010\tablefootmark{4}		&	"							&	$2.0\pm0.1$			&	$144.9\pm3.6$						\\
2010\tablefootmark{5}		&	"							&	$1.6\pm0.1$			&	$197.3\pm3.0$						\\
\hline
		\end{tabular}\\
	\tablefoottext{1}{\citet{Whitney_1993}},\tablefoottext{2}{\citet{Szeifert_2010}},\tablefoottext{3}{ExPo: Region 1},\tablefoottext{4}{ExPo: Region 2}
	\tablefoottext{5}{ExPo: Region 3}
	\label{tab:tab2}
\end{table}
\section{Discussion}
Z Cma was observed with ExPo during its last minimum. According to the current model (see Fig.~\ref{fig:zcma_fig1}), this implies that the primary star is obscured from sight, i.e., 
the medium is optically thick in the line of sight between the primary and us. At this stage, the primary contributes about 20\% of the total flux at optical wavelengths, and all its 
light is linearly polarized by the dust cocoon which surrounds it.

The process of removing the instrumental polarization, as explained in Sect. 2, will produce a systematic error in our measurements in case the central star(s) is polarized.
We can not disentangle the instrumental polarization from the intrinsic polarization associated to Z CMa. The dependence of the instrumental polarization
with the telescope pointing position prevents us from using the diskless star as an instrumental polarization calibrator: Z CMa and G 9143 were observed at different nights
and telescope orientations. A characterization of the instrumental polarization would require a detailed model of the telescope $+$ instrument system
\citep[see for example][]{Witzel_2011}, and this is beyond of the scope of this paper.
On the other hand, in the case of Z CMa the flux from the primary contributes up to $20\%$ of the total flux at optical wavelengths, and is polarized
\citep{Whitney_1993,Szeifert_2010}. The exact value of this polarization is not well known, with values ranging from $1.4\%$ to $2.6\%$. Furthermore, 
this value is expected to change with the state of the dust cocoon
surrounding the primary. As an extra source of error, the seeing severely affects the polarization degree, by reducing its true value when the seeing increases.
Because of all this our results must be taken as lower limits of the true intrinsic degree of polarization
of the three different regions described below.

\subsection{Region 1}
The polarization angle ($P_\mathrm{\Theta}$) inside this area is almost perpendicular to the direction of the blue-shifted side of the parsec-sized jet, as it is shown in the left 
image of Fig.~\ref{fig:zcma_fig5}. Moreover, this region follows the trajectory of the jet, as it can be seen on the right image of Fig.~\ref{fig:zcma_fig5}. 
\citet{Whitney_Model_1993} show that the polarization pattern that we measure in this region, with $P_\mathrm{\Theta}$ perpendicular to the jet direction, is indeed the expected pattern 
when observing an infalling dust envelope with a cylindrical cavity carved out by a well-collimated jet. 

Previous spectropolarimetric studies at optical wavelengths \citep{Whitney_1993,Szeifert_2010} have reported a polarization angle perpendicular to the direction of the parsec-sized jet. 
The average $P_\mathrm{\Theta}$ that we measure in this region are in good agreement with these previous observations, as it is shown in Table~\ref{tab:tab2}.
These results clearly suggest that this polarized feature is produced by scattering on the walls of the cavity carved out in the dusty envelope by the large jet.
We therefore point to the large jet associated to the primary as the origin of the polarization pattern observed in Region 1 of our images.

\subsection{Region 2}

This region shows the highest polarization degree in our Z CMa images. The trajectory of the micro-jet described by \citet{Whelan_2010}, $235^{\circ}$,
is almost perpendicular to the polarization angle in this region: $144.9^{\circ}\pm3.6^{\circ}$. As with the polarized feature in Region 1, the polarized feature
that we observe here is consistent with the polarization pattern predicted for an infalling envelope with empty cavities caused by a collimated outflow
\citep{Whitney_Model_1993}. This strongly points towards single scattering in the walls of a cavity carved out by this so-called micro-jet as the source of this polarized feature.

Our observations independently confirm the result from \citet{Whelan_2010}, and indicate that the micro-jet actually extends to a distance 
at least ten times larger than reported by these authors.

\subsection{Region 3}
The polarization angle of this area is very distinct from the polarization angles measured in previous observations. The average value of $P$ inside this area is
$P =  1.6\pm0.1\%$ (3 sigmas above the background polarization), and the location of this region in our image is 
not close to any of the two jets associated to Z CMa. Therefore, we argue that this polarized feature is not related to any the known jets. To explain its origin, we consider
here the possibility of a hole in the dust cocoon surrounding the primary star pointing to the West. Some of the outbursts of Z CMa, as well as the spectropolarimetric
measurements of this system, can be explained \citep{Whitney_1993,Szeifert_2010} by the formation of holes in the dust cocoon surrounding the primary: in case there
is a hole in the dust cocoon in our line of sight, then an increase in the amount of light from the primary as well as a decrease on its polarization degree is expected. 

To understand the impact of a hole in the optically thick dust cocoon surrounding the Herbig Be star, we model, using the MCMax radiative transfer code \citep{Min_2009},
a pencil beam of light emerging from the dust cocoon which is scattered towards the observer by a surrounding infalling molecular cloud.

We are aware of the limitations of this simulation when comparing to our observations. Since we cannot quantify
the instrumental polarization with enough accuracy, a direct comparison between our observations and the model can be
misleading. Thus, the primary goal of this modeling approach is to test wether a hole in the dust cocoon can produce a similar
polarized feature to what we observe in Region 3 rather than to provide detailed quantitative values for the parameters of the
surrounding material.

\subsubsection{Model Setup} 
We have simulated a binary system placed at a distance of 1150 pc (a value which agrees to the approximate distance to Z CMa), comprising an FU Ori star
and a Herbig Be star. The basic parameters of the FU Ori star are taken from \citet{Ancker_2004}, resulting in an star of $M = 3 M_{\sun}, L = 27.5 L_{\sun}$.
The luminosity of the embedded Herbig star is very poorly constrained in the literature, with values ranging from $L = 3000 L_{\sun}$ \citep{Ancker_2004} to 
$L = 310000 L_{\sun}$ \citep{Hartmann_1989}. For our model, we consider a Herbig Be star of
$M = 13 M_{\sun}, L = 55000 L_{\sun}$, which is in good agreement with the values obtained by \citet{Alonso_2009}, surrounded by an optically thick dust cocoon.
The binary system is surrounded by a large infalling molecular envelope. The dust cocoon surrounding the Herbig Be star is modelled as an optically thick spherical
shell with a hole with an opening angle ($\phi$) varying from $\phi = 5^{\circ}$  to $\phi = 10^{\circ}$. This cocoon is 50 AU in radius, starting at a distance of 20
AUs from the primary, which is the dust sublimation radius for this star. In this computation, three different orientations (with respect to the line of sight) of the
hole in the dust cocoon were tested: $+45^{\circ}$ (producing forward scattering), $-45^{\circ}$ (producing backward scattering), and $90^{\circ}$. The infalling
envelope is modelled as an optically thin spherical shell with a density distribution described by \citet{Dominik_Dullemond_2008}, based on the work of
\citet{Ulrich_1976} and \citet{Terebey_1984}. In this model, the infalling envelope is made from the leftovers from the collapse of a rotating cloud. The angle between
a point on the envelope, the center of the envelope, and the rotational axis defines de polar coordinate $\theta$. According to this model, the gas density $\rho$
at a polar angle $\mu = \mathrm{cos}$ $\theta$ and a distance to the center $r$ is given by:
\begin{equation} 
\rho(r,\theta) =  \frac{\dot{M}}{4\pi\sqrt{GMr^{3}}} \left( 1 + \frac{\mu}{\mu_{0}} \right)^{-1/2} \left(\frac{\mu}{\mu_{0}} + \frac{2\mu_{0}^{2}r_\mathrm{centr}}{r} \right)^{-1},
\label{eq:eq5}
\end{equation} 
where $\mu_{0}$ is the solution of the equation
\begin{equation} 
\mu_{0}^{2} = 1 - \frac{r}{r_\mathrm{centr}}\left(1 - \frac{\mu}{\mu_{0}} \right).
\label{eq:eq6}
\end{equation} 
$\dot{M}$ represents the infall rate and $r_\mathrm{centr}$ is the centrifugal radius, which here is fixed at a distance of 200 AU, as in the model of
\citet{Dominik_Dullemond_2008}. Four different values of $\dot{M}$ $(1,2,4,8\times10^{-7}$ $M_{\sun}$ $\mathrm{yr}^{-1}$) were computed in our simulations.
To easily distinguish between the different models we use the following notation: ``Model 1(5b)" refers to the model with 
$\dot{M} = 1\cdot10^{-7} M_{\sun}\mathrm{yr}^{-1}$, $\phi = 5$, backward scattering; ``Model 4(10f)" refers to the model with
$\dot{M} = 4\cdot10^{-7} M_{\sun}\mathrm{yr}^{-1}$, $\phi = 10$, forward scattering, ``Model 4(10r)" refers to the model with
$\dot{M} = 4\cdot10^{-7} M_{\sun}\mathrm{yr}^{-1}$, $\phi = 10$, $90^{\circ}$ scattering, 
and so on. The composition of the dust in the infalling envelope is described
by \citet{Min_2011}, which is based on the Solar composition inferred by \citet{Grevesse_1998}. The infalling envelope is composed by, in mass, $58\%$ silicates,
$18\%$ iron sulphide and $24\%$ amorphous carbon. The dust size distribution follows the standard MRN size distribution \citep{Mathis_1977}. In this distribution
the number density of the grains is equated by the following power-law
\begin{equation} 
n(a) \mathrm{d}a \propto a^{-3.5} \mathrm{d}a,
\label{eq:eq7} 
\end{equation} 
where $a$ is the radius of the grains which ranges from 5 nm to 250 nm. The shape of the dust grains follows the distribution of hollow spheres
\citep[DHS,][]{Min_2005}, assuming an ``irregularity parameter" $f_\mathrm{max} = 0.8$. 

The radiative transfer images of the Z CMa system modelled by MCMax are then convolved with the modelled Point Spread Function (PSF) of ExPo from
\citet{Min_2012}. The PSF is modelled with an average seeing of 1.4" (as in our observations of Z CMa) and includes contributions from photon
noise, readout noise, and instrumental polarization. To reproduce the ExPo observations, the exposure time used for these PSFs is 0.028 seconds. 
The resulting PSF is then a speckle pattern, similar to what we obtain with ExPo observations.
To simulate the effect of the instrument and the data processing, we proceed as follows. We first produce a set of 100 statically independent,
noiseless, PSFs. The images produced by the MCMax code
are then convolved with one simulated PSF. We then compute the photon noise associated to this new image. This noise is re-scaled to match
the amplitude of the photon noise associated to a PSF with exposure time of 2.8 seconds. We then add this noise to the simulated image, 
and we repeat this process with all the PSFs. At the end of the process we have a set of 100 different images, each one with the photon noise
associated to an image of 2.8 seconds exposure time (or 100 images taken with 0.028 seconds of exposure time). This is roughly equivalent
to compute 10000 independent ExPo PSFs, but is much faster in terms of computing time. Instrumental polarization is added before the images
are reduced with the ExPo pipeline described in Sect. 2.2 of this paper and \citet{Canovas_2011}. The resulting images produced after the reduction
include then both instrumental and data-processing effects, such as instrumental polarization or un-perfect alignment. After reducing these
simulated images, the averaged PSF has a FWHM of  $\sim1.2"$, very similar to what we obtain after reducing our data. Both stars appear unresolved in the
simulation, as in our observations. Finally, we reduce the noise in these images by convolving them with a gaussian kernel of 5 pixels FWHM,
as we do with our observations.

To compare with the real images and decide which model is the more realistic, the mean value of the polarization degree in an area with the 
same size and shape as Region 2 in Fig.~\ref{fig:zcma_fig5} was computed. We have also computed the background polarization (BP) as the mean
of the degree of polarization inside of half a ring with inner radius $r_\mathrm{in} = 4.1"$ and outer radius $r_\mathrm{out} = 6.1"$ centered on the binary position.
By doing this, we do not include the contribution of the polarized feature caused by the hole on the cocoon. This quantity systematically increases when
increasing the the mass infalling ratio, and it is practically insensitive to the size of the hole in the dust cocoon, as expected from our approach.
Table ~\ref{tab:tab4} shows the average $P$ of the final simulated-image and the background polarization for each one of the models described above.

\begin{figure} 
  \center 
   \includegraphics[width = 1.\linewidth,trim = 80 80 80 95]{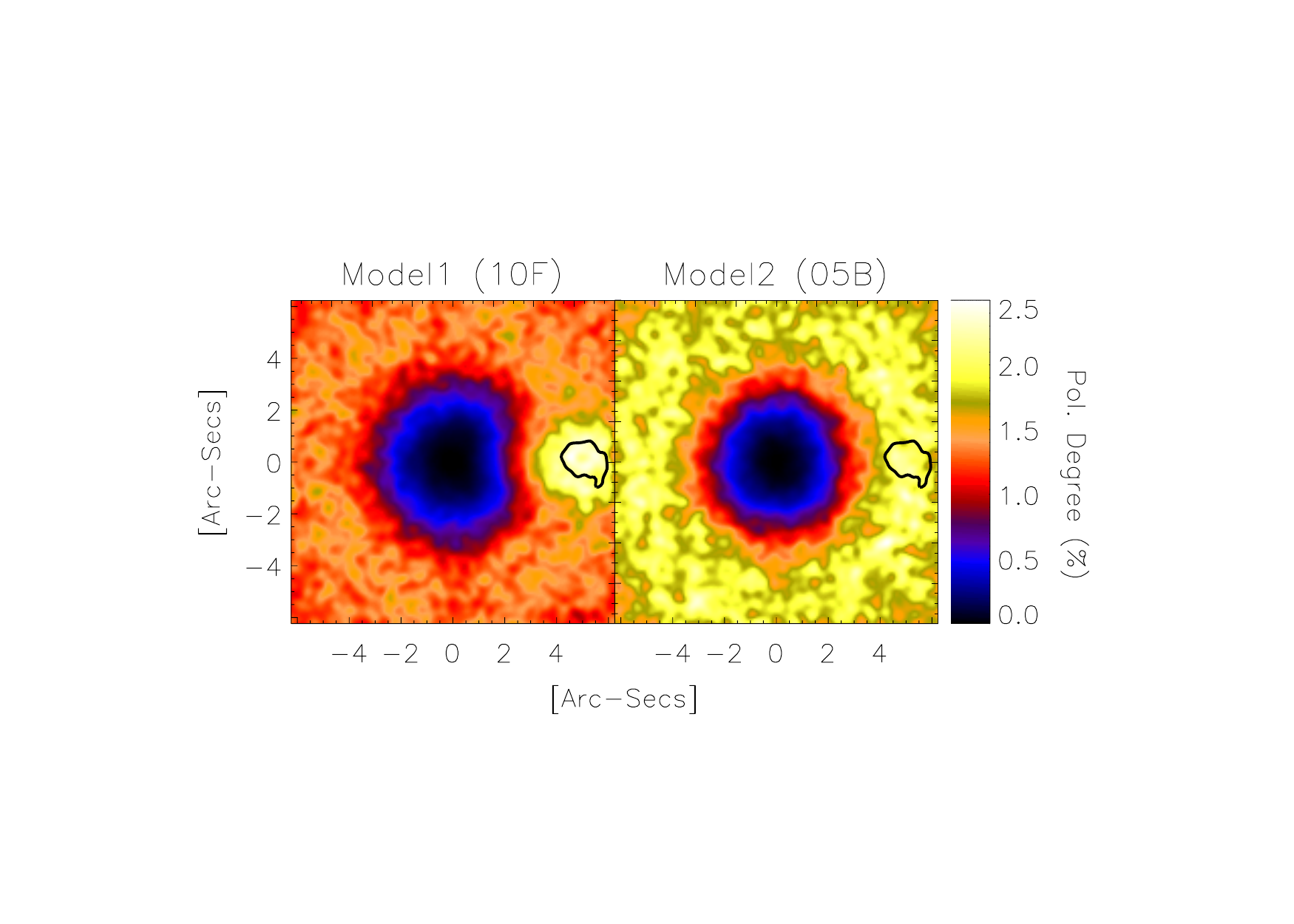} 
   \caption{Two models that reproduce the degree of polarization in an area equal to Region 3 in Fig.~\ref{fig:zcma_fig5}. The model on the right produces an excess
   of background polarization, which translates into a bright ring around the (unpolarized) stars. The black lines are contouring an area equal to Region 3  (see Fig.~\ref{fig:zcma_fig5}).}
  \label{fig:zcma_fig6} 
\end{figure}

\begin{table*}[h] 
\centering
\caption{Mean values of $P$ for the different models computed.} 
\begin{tabular}{c c c c c c c c c c c c c} 

&       \multicolumn{4}{c}{$-45^{\circ}$}\vline     &       \multicolumn{4}{c}{$45^{\circ}$  }\vline	&	\multicolumn{4}{c}{$90^{\circ}$ } \\ 

\cline{2-13}

$\dot{M}$							&$P_{\phi = 5}$&	BP		&$P_{\phi = 10}$	&BP		&$P_{\phi = 5}$&	BP		&$P_{\phi = 10}$	&	BP	&$P_{\phi = 5}$	&	BP	&	$P_{\phi = 10}$&BP\\
$[10^{-7}M_{\sun}\mathrm{yr}^{-1}]$		& $[\%]$		&	$[\%]$	&$[\%]$			&$[\%]$	&$[\%]$		&	$[\%]$	&$[\%]$			&	$[\%]$&$[\%]$			&$[\%]$	&$[\%]$			&$[\%]$\\
\hline 
1											&	1.1&1.0&1.3&1.0	&1.2&1.0&1.7&1.0&1.3&1.0&2.4&1.0			\\
2											&	1.6&1.3&2.1&1.3	&1.9&1.3&2.8&1.3&2.3&1.3&4.2&1.3			\\
4											&	2.5&1.9&3.4&1.9&2.9&1.9&4.0&1.9&3.4&1.9&6.6&1.8			\\
8											&	3.2&2.2&4.2&2.3	&3.6&2.3&4.3&2.3&4.7&2.2&8.3&2.2			\\
\hline 
                \end{tabular} 
        \tablefoot{$\phi$ stands for the opening angle (in \textit{degrees}) of the hole in the dust cocoon. BP stands for background polarization, measured in a
        half ring, with $r_{in} = 4.1"$ from the center, and $r_{out} = 6.1"$. The average (simulated) PSF used for these simulations 
        has a FWHM of $1.2"$ after been realigned with the ExPo pipeline.}
\label{tab:tab4} 
\end{table*} 
We find that there are two models that produce an average polarization value similar to what we measure
in Region 3 of our image. These two models are ``Model 1 (10f)", which produces $P = 1.7\%$, and ``Model 2 (5b)", which produces $P = 1.6\%$.
However, the model with the highest accretion rate fails when trying to reproduce the polarization background ($BP = 1.3\%$).
Because of the highest accretion rate,
there is more dust contributing to the polarized background. This is clearly illustrated in Fig~\ref{fig:zcma_fig6}. The bright, yellow blob that appears on the
left image roughly resembles the Region 2 of Fig.~\ref{fig:zcma_fig5}.

Given the so many unknowns of the Z CMa system, it is not possible to constrain all its details. Because all the assumptions
that we had to make during our modeling (i.e., hole geometry, dust shell geometry, dust composition, etc), we must consider our results
as a very first indication. However, and within the uncertainties of our model, we obtain two results.
First, the mass infall rate must be lower than $\dot{M} = 2\cdot10^{-7} M_{\sun} \mathrm{yr}^{-1}$ to fit our model.
Second, and perhaps the most important, a new hole on the dust cocoon surrounding the primary can explain the
polarized feature observed in Region 3 of our images.

\section{Conclusions}

Our measurements, obtained with ExPo after the 2008 outburst, show three different features in polarization. Two of them, labeled here as ``Region 2'' and ``Region 3'' 
have not been reported before. This is due to the high polarimetric sensitivity of ExPo, and the different nature of our data, which employ imaging polarimetry
rather than spectropolarimetry. We interpret our data as new evidence of the presence of holes in the dust cocoon that surrounds the primary star:
our simulation about the effects of one hole in the dust cocoon can reproduce the polarized feature observed in Region 3.

The polarized feature labeled here as Region 1 in Fig.~\ref{fig:zcma_fig5} shows an average polarization angle of $P_\mathrm{\Theta} = 155.4^{\circ}\pm3.9^{\circ}$, which
is perpendicular to the direction of the large jet and it is very similar to the value found by \citet{Whitney_1993} and \citet{Szeifert_2010}. The orientation of
$P_\mathrm{\Theta}$ indicates that this polarized feature is consistent with scattering on the walls of a cavity carved out by the large jet, which trajectory is perpendicular
to $P_\mathrm{\Theta}$. \citet{Benisty_2010} show evidence of a tilted accretion disk around the primary by analyzing the $\mathrm{Br}_{\gamma}$ emission line produced 
on the hot gas around this star. If this is the case, then this disk must be inside of the dust cocoon which surrounds this star, to explain all the previous evidence of a  dust 
cocoon around the primary.

The average polarization angle in Region 2 is perpendicular to the trajectory of the micro-jet. This strongly suggests that
this feature is caused by single scattering originated in the walls of a cavity carved out by this jet. The results from \citet{Whelan_2010}
indicate that this jet extends up to a distance of 0.5" from the FU Ori star, while the polarized feature that we observe appears at a distance of $\approx 5"$
from the FU Ori star. This indicates that this micro-jet actually extends to sizes much larger than reported before.

Our analysis suggests that the polarized feature that we detect in Region 3 in Fig.~\ref{fig:zcma_fig5} is caused by a hole in the dust cocoon surrounding the primary star.
The idea of a hole in the dust cocoon is not new and it has been used to explain the variability and polarimetric properties of this system. However, in previous discussions,
the hole in the cocoon was assumed to be in our line of sight to explain some of the outbursts of Z CMa. To fit our observations, a hole with a different orientation must be
used: a hole in one side of the cocoon can produce a pencil beam of light (as a lighthouse) that will be scattered to our line of sight, producing a polarized signature similar
to what we observe in Region 3. To simulate this, we have run different models with different hole sizes, orientations, and different mass infall rates for the surrounding infalling
envelope. We conclude that a hole in the dust cocoon can produce a polarized beam of light that explains the polarized feature observed in Region 3.
According to our calculations and models, the mass infall rate of the infalling envelope cannot be higher than $\dot{M} = 2\cdot10^{-7}M_{\sun}\mathrm{yr}^{-1}$.
Given the error bars on the luminosity of the primary star \citep[see, for instance,][]{Monnier_2005}, together with the many assumptions made in our model,
we take this value as a first approximation. The derived infall rate scales inversely with the luminosity of the system; so in case we are overestimating the luminosity
of the Herbig Be star, we are underestimating the infall rate, and vice-versa.

Region 3 requires more observations for a proper analysis of our results. The nature of the dust cocoon around the primary remains a puzzle, with many
open questions about it such as its size or grain composition. The source of these holes and/or variability in the dust cocoon remains still unknown, but one
possible explanation might be related to the nature of this system: given the small separation between the primary and the secondary, their gravitational interaction
might affect the dust cocoon around the primary \citep[as an example of the effect of binarity over circumstellar disk sizes, see ][]{Artymowicz_1994ApJ}.

Z CMa is a system with an extremely high variability. However, the different nature of the data discussed by previous
authors \citep{Whitney_1993,Szeifert_2010} and ours leaves no room to discuss this variability in terms of the observations presented here
and the previous ones. On the other hand, the keplerian timescale \footnote{The keplerian timescale
indicates the orbital period around a central object of mass $M_{\star}$. It can be approximated as $\tau \sim \frac{2\pi}{\sqrt{GM_{\star}}} \left( \frac{R}{1 - e}\right)^{2/3}$,
where $R$ is the orbital radius, $e$ is the eccentricity of the orbit and $G$ is the general constant of gravity.} of a dust particle orbiting a $M = 13 M_{\sun}$ star at a
distance equal to the sublimation radius (20 AUs in this case) is approximately 24 years, and about 100 years when the distance to the star is 50 AUs. Therefore,
one might expect to measure variability caused by the movement of the dust cocoon around the primary on timescales of years to decades.

A characterization of the dust properties of this system will help to constrain the huge error bars associated to the luminosity of the primary star, helping to
solve part of the unknowns of this complex system. Furthermore, the determination of the infall rate could help to explain mechanisms that produce the extreme
accretion rates measured for the FU Ori star \citep[see][]{Hartmann_1996}.

Z CMa is a very interesting system to perform imaging polarimetry studies. If the dust cocoon around the primary is rapidly modifying its geometry, then new holes (i.e., new polarized
features) are expected to appear. According to our results, this could happen on timescales of the order of a few years. Future observations of this system with imaging
polarimeters such as ExPo or SPHERE will provide crucial information helping to understand the environment of Z CMa.

\begin{acknowledgements}
	The authors thank the anonymous referee for providing very constructive comments, 
	which significantly helped to improve the paper. We are also grateful to the staff at
	the William Herschel Telescope for their help during the during the observations.
	H.C.C. acknowledges
	support from Millenium Science Initiative, Chilean Ministry of Economy,
	Nucleus P10-022-F.
\end{acknowledgements}

\bibliographystyle{aa.bst}	
\bibliography{biblio}		
\end{document}